\documentclass[aps,prl,reprint,superscriptaddress,amsmath,amssymb,floatfix,longbibliography]{revtex4-1}
\usepackage{graphicx}
\usepackage{bm}
\usepackage{color}
\usepackage{lipsum}
\usepackage{xcolor}

\begin{document}

\title{Heating and Cooling of Quantum Gas by Eigenstate Joule Expansion}

\author{Jae Dong Noh}
\affiliation{Department of Physics, University of Seoul, Seoul 02504,
Korea}
\author{Eiki Iyoda} 
\affiliation{Department of Applied Physics, The University of Tokyo, 7-3-1
Hongo, Bunkyo-ku, Tokyo 113-8656, Japan} 
\author{Takahiro Sagawa} 
\affiliation{Department of Applied Physics, The University of Tokyo, 7-3-1
Hongo, Bunkyo-ku, Tokyo 113-8656, Japan} 

\date{\today}

\begin{abstract}
We investigate the Joule expansion of an interacting quantum gas 
in an energy eigenstate. 
The Joule expansion occurs when two subsystems of
different particle density are allowed to exchange particles.
We demonstrate numerically that the subsystems in their energy eigenstates
evolves unitarily into the global equilibrium state in accordance with the 
eigenstate thermalization hypothesis. 
We find that the quantum gas changes its temperature after the Joule
expansion with a characteristic inversion temperature $T_{\rm I}$.
The gas cools down~(heats up) when the initial temperature 
is higher~(lower) than $T_{\rm I}$, implying that $T_{\rm I}$ is a stable fixed 
point, which is contrasted to the behavior of classical gases.
Our work exemplifies that transport phenomena can be studied at the level 
of energy eigenstates.
\end{abstract}
\maketitle

{\it Introduction --- }
Statistical mechanics postulates that an isolated quantum system 
in thermal equilibrium is represented by the completely mixed state in the
microcanonical energy shell.
It has been a puzzling question whether statistical mechanics is compatible
with unitary dynamics of quantum mechanics which does not allow a transition 
of a pure state to a mixed state.
Recent studies have revealed that this puzzle can be settled in view of
quantum ergodicity~\cite{{DAlessio:2016gr}}.
A quantum mechanical system in a pure state can be
thermal by itself. That is, the system, if quantum chaotic, plays a role of an
equilibrium heat bath for its subsystem as if it were in the equilibrium
mixed state. In fact, the
eigenstate thermalization hypothesis~(ETH) asserts that all the energy
eigenstates are thermal for a broad class of
non-integrable quantum 
systems~\cite{{Srednicki:1994dl},{Deutsch:1991ju},{Kim:2014kw},{Yoshizawa:2018js},{Rigol:2009ew},{vonNeumann:2010jl},{Beugeling:2014ci}}.

The ETH has been tested numerically in various discrete lattice systems.
Those studies confirm that the expectation value of local observables 
in the energy eigenstate is consistent with the statistical mechanics
prediction~\cite{{Jensen:1985vm},{Rigol:2008bf},{Kim:2014kw},{Yoshizawa:2018js}}. 
They also confirm that quantum systems thermalize
after a quench, a sudden change in the Hamiltonian, following the ETH 
prediction~\cite{{Rigol:2009ew}}. The thermalization of isolated quantum
systems has also been studied experimentally using ultracold
atoms~\cite{{Kinoshita:2006bg},{Tang:2018dq},{Trotzky:2012iu},{Langen:2015do},{Kaufman:2016jm},{Kim:2018jr}}
and superconducting qubits~\cite{Neill:2016ff}. 
The ETH is now recognized as a paradigm of statistical mechanics 
for pure quantum systems with a few notable exceptions such as the
integrable systems~\cite{Biroli:2010fz}, the many-body localization
systems~\cite{Imbrie:2016ib}, systems with many-body quantum
scars~\cite{Turner:2018iz,Shiraishi:2017ek,*Mondaini:2018jo,*Shiraishi:2018gc}.

Besides a single isolated system, 
it is also interesting to ask how two quantum systems thermalize 
in the presence of a thermal contact. 
Ponomarev {\it et al.} demonstrated numerically 
the thermalization of two systems which exchange the 
energy~\cite{{Ponomarev:2011bi}}. 
A thermal contact may also allow the exchange of a 
globally conserved entity such as the particle number. Yet, the quantum 
thermalization under such a contact has been studied rarely.

\begin{figure}
\includegraphics[width=\columnwidth]{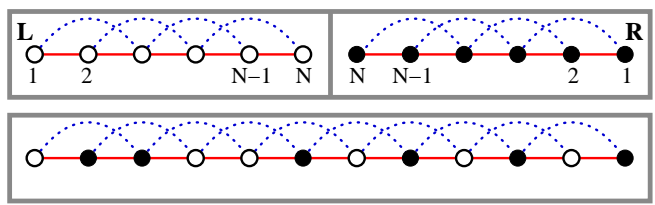}
\caption{Joule expansion of a quantum gas. Filled~(empty)
circles represent the occupied~(empty) lattice sites. Also drawn is 
the lattice structure with site indices for the Hamiltonian.}\label{fig1}
\end{figure}

The Joule expansion is a representative irreversible process taking place 
under the general contact~\cite{{Reif:1965uf},{Blundell:2006we}}.
Suppose that a gas is confined in a compartment of an isolated
container~(see Fig.~\ref{fig1}). When the dividing wall is removed or a
contact opens, the gas expands irreversibly and reaches the homogeneous 
equilibrium state. The Joule expansion in the classical regime is well
understood. For instance, the Van der Waals gases cool down upon expansion 
because gas particles loose the kinetic energy gaining the attractive 
interaction potential energy~\cite{Blundell:2006we}. 
A mean field study with the Lennard-Jones potential showed that the classical 
gases can heat up if the initial temperature is high above a threshold
$T_{\rm I}$, called the inversion temperature~\cite{Goussard:1993iy}. 
The short range repulsion between particles is responsible for the inversion
temperature.

In this Letter, we investigate the thermalization of two quantum 
systems coupled by an interaction Hamiltonian which allows the exchange of 
particles as well as the energy.
Both subsystems are in their respective energy eigenstates, and evolve
unitarily into a steady state after the interaction turns on. 
With this setup, we can study the Joule expansion of an interacting quantum gas 
in an energy eigenstate~\cite{*[{The Joule expansion of noninteracting quantum 
gas was studied in }] [{.}] {Camalet:2008fg}}. 
We demonstrate that there exists an inversion temperature $T_{\rm I}$: 
The quantum gas cools down~(heats up) when the initial temperature is
above~(below) $T_{\rm I}$, which makes the inversion temperature stable.
This is in sharp contrast to the behavior of the classical
gases which heats up when the initial temperature is
above an inversion temperature, 
which suggests that the inversion temperature results from 
a many-body correlation effect in the quantum case. 
The Joule expansion can be realized experimentally using ultracold 
gases~\cite{Schneider:2012ke}. 
We anticipate that our result will be relevant for 
controlling temperature in such experimental setups.

{\it Model ---}
For concreteness, 
we present our work in the context of a spin chain system. 
We consider the two identical spin-1/2 XXZ chains of length $N$, referred 
to as L and R, with nearest and next nearest neighbor interactions~(see
Fig.~\ref{fig1}).
The Hamiltonian of each chain $\alpha$ = L, R reads
\begin{equation}\label{Hchain}
\hat{H}_\alpha = \frac{1}{1+\lambda} \left( 
\sum_{i=1}^{N-1} h(\hat{\sigma}_{\alpha,i} , \hat{\sigma}_{\alpha,i+1}) +
\lambda \sum_{i=1}^{N-2} h(\hat{\sigma}_{\alpha,i} , \hat{\sigma}_{\alpha,i+2})
\right) 
\end{equation}
with the two-body interaction Hamiltonian 
\begin{equation}
h(\hat{\sigma}_i , \hat{\sigma}_j) = -\frac{J}{2} 
\left( \hat{\sigma}_{i}^x \hat{\sigma}_{j}^x + \hat{\sigma}_{i}^y \hat{\sigma}_{j}^y + \Delta
\hat{\sigma}_{i}^z \hat{\sigma}_{j}^z \right) .
\end{equation}
Here, $\hat{\sigma}_{\alpha,i}$ denotes the Pauli matrix for a spin at
site $i~(=1,\cdots,N)$ in the chain $\alpha$. 
The model includes a few parameters: $J>0$ sets the 
scale of energy, $\Delta$ is the anisotropy parameter,
and $\lambda$ represents the relative strength of the next nearest neighbor
interactions. 
With nonzero $\lambda$, the system is known to satisfy the 
ETH~\cite{Rigol:2009ew,Yoshizawa:2018js}.  We will set $J$ to unity. 
As a thermal contact, we adopt an interaction Hamiltonian
\begin{equation}
\begin{split}
\hat{H}_{\rm int} = & \frac{1}{1+\lambda} [ 
h(\hat{\sigma}_{{\rm L},N},\hat{\sigma}_{{\rm R},N}) +  \\
 &  \lambda h(\hat{\sigma}_{{\rm L},N-1},\hat{\sigma}_{{\rm R},N}) + \lambda
h(\hat{\sigma}_{{\rm L},N},\hat{\sigma}_{{\rm R},N-1}) ] .
\end{split}
\end{equation}
With this choice, the total
Hamiltonian $\hat{H}_{\rm tot} = \hat{H}_{\rm L} + \hat{H}_{\rm R} + \hat{H}_{\rm int}$ becomes
that of the XXZ chain of $2N$ spins. 
The conclusion is not altered with a choice of different coupling
constants in $\hat{H}_{\rm int}$. In the numerical study, the parameter
values are $\Delta = 1/2$ and $\lambda=1$ unless stated otherwise.

The spin-1/2 chain system is equivalent to a hardcore boson
system~\cite{{Cazalilla:2011dm}} by identifying a site where the $z$ 
component of spin is up as an occupied site by a bosonic particle. 
Each site can be occupied by at most a single particle.
In the context of the boson system, the
coupling in the $x$ and $y$ directions corresponds to the kinetic energy
term and the coupling in the $z$ direction corresponds to the
attractive~$(\Delta >0)$ or repulsive $(\Delta < 0)$ interaction between
particles. 

\begin{figure}[th]
\includegraphics*[width=\columnwidth]{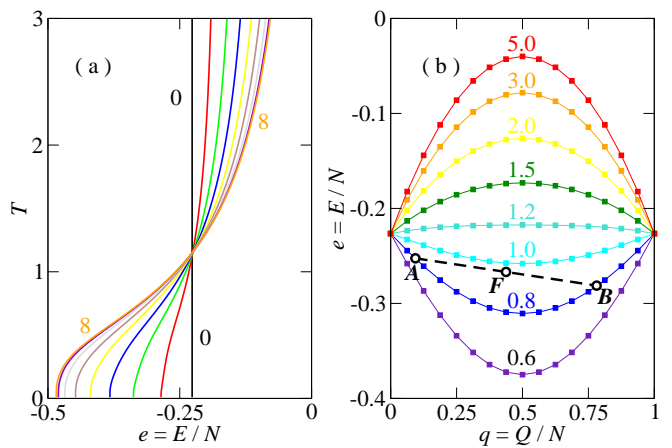}
\caption{(a) Energy density-temperature relation at each $Q~(=0,\cdots,8)$ 
sector.
(b) Isothermal curves at the specified temperatures.
These are obtained by diagonalizing numerically the Hamiltonian
\eqref{Hchain} with $N=16$, $\Delta = 1/2$, and $\lambda=1$.}
\label{fig2}
\end{figure}

Before addressing the thermalization of the total system, we
summarize the thermal property of the subsystem.
The Hamiltonian \eqref{Hchain} commutes with 
$\hat{Q}_\alpha = \sum_{i=1}^N \frac{(1+\hat\sigma_{\alpha,i}^z)}{2}$ 
that counts the number of up spins or particles in the subsystem $\alpha$.
Thus, one may consider the subspace of the Hilbert space in which 
$Q~(=0,1,\cdots,N)$, the eigenvalue of $\hat{Q}_\alpha$, is fixed, separately.
It is called the $Q$ sector. Due to the particle-hole symmetry, the $Q$
sector is equivalent to the $(N-Q)$ sector.
Let $|Q,n\rangle$ with $n = 1 , \cdots, \binom{N}{Q}$ be the eigenstate 
with the $n$th lowest energy eigenvalue $E_{Q,n}$ in the
$Q$ sector. 
With $\lambda\neq 0$, the system is thermal so that 
an energy eigenstate $|Q,n\rangle$ can be assigned
to a temperature $T_{Q,n} = 1/\beta_{Q,n}$ from the 
relation~\cite{Santos:2011ip}
\begin{equation}\label{T_E}
E_{Q,n} = \sum_{m=1}^{\binom{N}{Q}} E_{Q,m} e^{- \beta_{Q,n} E_{Q,m}} / Z_Q
\end{equation}
with the partition function $Z_Q = \sum_m e^{-\beta_{Q,n} E_{Q,m}}$. The
Boltzmann constant is set to be unity. Figure~\ref{fig2}(a) presents the
energy-temperature relation in each $Q$ sector.
Also shown in Fig.~\ref{fig2}(b) are the isothermal curves. 
The isotherms have a positive or
negative curvature depending on the temperature. 
The curvature change leads to an intriguing phenomenon, which will be
discussed later.

{\it Thermalization ---}
Suppose that the total system is prepared to be in a product state
\begin{equation}\label{Psi_ini}
|\Psi(0) \rangle =  |Q_{\rm L}^0,n_{\rm L}^0\rangle \otimes | Q_{\rm
R}^0,n_{\rm R}^0\rangle ,
\end{equation}
where $|Q_{\rm L}^0,n_{\rm L}^0\rangle$ and $|Q_{\rm R}^0,n_{\rm R}^0\rangle$ are the eigenstates of 
$\hat{H}_{\rm L}$ and $\hat{H}_{\rm R}$, respectively. 
The interaction Hamiltonian does not
commute with $\hat{H}_\alpha$ and $\hat{Q}_\alpha$. 
Thus, $\hat{H}_{\rm int}$ acts as 
a thermal contact allowing the flows of the energy and the
particle.
We investigate how the system evolves into the global equilibrium state
via the unitary time evolution
$|\Psi(t) \rangle = e^{-i t \hat{H}_{\rm tot}} 
|\Psi(0)\rangle$ with $\hbar=1$. The time evolution is simulated
numerically~\cite{Suzuki:1985fz}~\footnote{See Supplemental Material for more details.}.

\begin{figure}[t]
\includegraphics*[width=\columnwidth]{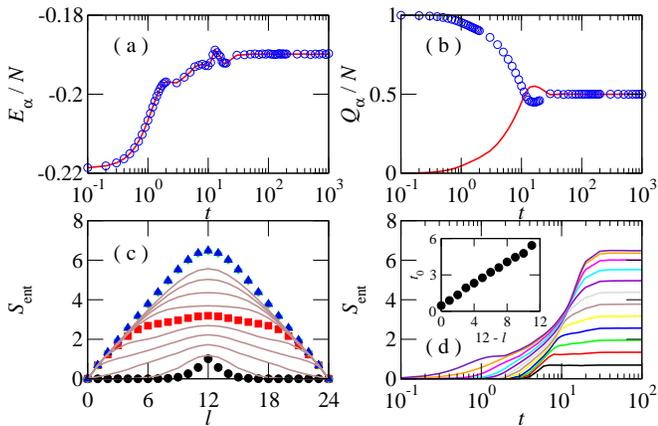}
\caption{Time evolution starting from the initial state $|\Psi(0) \rangle = 
|Q_{\rm L}=0,n_{\rm L}=1\rangle \otimes |Q_{\rm R}=N,n_{\rm R}=1\rangle$
with $N=12$, $\Delta=1/2$, and $\lambda=1$.
(a) and (b) Energy densities and particle densities for the subsystem 
${\rm L}$~(line) and ${\rm R}$~(symbol).
(c) The entanglement entropy at $t=10^0$~(circle), $10^1$~(square),
$10^2$~(diamond), and $10^3$~(triangle). Data at intermediate moments are
drawn with lines. (d) Temporal behavior of the entanglement entropy
$S_{\rm ent}(l,t)$ at each value of $l=0,1,\cdots,12$ from bottom to top. Inset
shows the plot of $t_0$ at which $S_{\rm ent}(l,t_0)=0.5$.}
\label{fig3}
\end{figure}

We performed the numerical analysis with the initial state where the
subsystem L is empty~($Q_{\rm L}^0=0$) and the subsystem ${\rm R}$ is fully
occupied~($Q_{\rm R}^0=N$).
The expectation values of the energy 
$E_\alpha(t) = \langle \Psi(t) | \hat{H}_\alpha | \Psi(t) \rangle$ and the particle
number $Q_\alpha(t) = \langle \Psi(t) | \hat{Q}_\alpha | \Psi(t) \rangle$ of each
subsystem are measured and shown in Fig.~\ref{fig3}.
After a transient period, the system reaches a
stationary state where the energy and the particle are distributed
uniformly. 

The quantum thermalization is accompanied by the growth of the entanglement
entropy~\cite{{Kim:2013fy},{Kaufman:2016jm},{Nahum:2017ef}}. 
We construct the reduced density operator $\hat{\rho}_l(t) = {\rm
Tr}_{\bar{l}} |\Psi(t)\rangle \langle \Psi(t)|$ for the $l$ leftmost spins
by taking the partial trace over the other spins, 
and evaluate the entanglement entropy $S_{\rm ent}(l,t) = -{\rm Tr}
\hat\rho_l(t) \ln \hat\rho_l(t)$~\cite{Note1}. 
The numerical data shown in Fig.~\ref{fig3}(c) 
reveal two distinct transient regimes. The entanglement entropy of the
initial product state is identically zero.
At short times, the entanglement develops near the contact~($l\simeq 12$). 
It propagates into the system until $S_{\rm ent}(l,t)$ deviates from zero at all
values of $l$.
Afterwards, the entanglement entropy grows and saturates to the stationary 
profile. Note that the faceted shape with rounded center, i.e., the Page
correction, is the typical entanglement entropy profile of a thermal 
system~\cite{{Page:1993ur},{Nakagawa:2018db}}.
The propagation and the growth are clearly seen
in Fig.~\ref{fig3}(d). We estimate the entanglement propagation speed by 
measuring the time $t_0$ at which $S_{\rm ent}(l,t_0) = 0.5$ as a function of
$l$. It scales 
linearly with the distance $(N-l)$ from the boundary. The linear dependence
reflects the Lieb-Robinson bound for the quantum information
propagation~\cite{{Lieb:1972wy},{Hastings:2006gv}}.

We investigate in more detail the property of the steady state.
In the presence of $\hat{H}_{\rm int}$, $|\Psi(0)\rangle$ is not the eigenstate of 
$\hat{H}_{\rm tot}$. The energy fluctuation is given by
\begin{equation}
\sigma_E^2 = \langle \hat{H}_{tot}^2 \rangle_0 -
\langle \hat{H}_{tot} \rangle_0^2 
           = \langle \hat{H}_{\rm int}^2 \rangle_0 -
\langle \hat{H}_{\rm int} \rangle_0^2  ,
\end{equation}
where $\langle \cdot \rangle_0 = \langle \Psi(0) | \cdot | \Psi(0)\rangle$.
Since $\hat{H}_{\rm int}$ is local, the variance is of the order of
$O(N^0)$. The total Hamiltonian satisfies the ETH and the initial state has
a finite energy fluctuation. Therefore, the system should thermalize to the
equilibrium state in the long time limit.

We characterize the equilibrium state with the probability distribution
$P_{\rm L}(Q,n)$ that the subsystem ${\rm L}$ is in the eigenstate $|Q_{\rm
L}=Q,n_{\rm L}=n\rangle$ 
of $\hat{H}_{\rm L}$. 
It is given by
\begin{equation}\label{P_def}
P_{\rm L}(Q,n) = \langle Q,n| \hat{\rho}_{\rm L}(t=\infty) | Q,n\rangle
\end{equation}
where $\hat{\rho}_{\rm L}(t) = {\rm Tr}_{\rm R} |\Psi(t)\rangle \langle
\Psi(t)|$ is the reduced density matrix of the subsystem ${\rm L}$. 
The total system is thermal and can be regarded as the equilibrium heat bath
for subsystems. Thus, the subsystem probability is given by
\begin{equation}\label{P_formal}
P_{\rm L}(Q,n) \propto \exp[S_{{\rm R}}(E_{\rm tot}- E_{Q,n},Q_{\rm tot}-Q)] ,
\end{equation}
where $E_{\rm tot} = \langle \hat{H}_{\rm tot}\rangle_0 \simeq
E_{Q_{\rm L}^0,n_{\rm L}^0}+E_{Q_{\rm R}^0,n_{\rm R}^0}$, $Q_{\rm tot} = Q_{{\rm L}}^0 +
Q_{\rm R}^0$, and $S_{\rm R}(E,Q)$ is the thermodynamic entropy of the subsystem ${\rm
R}$. Here, we assume the weak coupling limit that the interaction energy is
negligible.

Let $\delta E = E_{Q,n} - \overline{E_{\rm L}}$ and $\delta Q = Q -
\overline{Q_{\rm L}}$ denote the deviations of the energy and the particle number
from their steady state values $\overline{E_{\rm L}}$ and
$\overline{Q_{\rm L}}$. In terms of $\delta E$ and $\delta Q$, we can approximate
\eqref{P_formal} as
\begin{equation}\label{P_exp}
\begin{split}
P_{\rm L}(Q,n) \propto  \exp [& -(\delta E - \mu \delta Q)/T 
 - a_{11} (\delta E)^2 - \\
& a_{12} (\delta E)(\delta Q) - a_{22} (\delta Q)^2 + \cdots] 
\end{split}
\end{equation}
with the temperature $T$ and the chemical potential $\mu$ defined as
$\frac{1}{T} = \left(\frac{\partial S_{\rm R}}{\partial E}\right)_Q$ and 
$\frac{\mu}{T} = -\left(\frac{\partial S_{\rm R}}{\partial Q}\right)_E$.   
We also keep the quadratic terms with the second order derivatives 
$a_{ij}$ of the thermodynamic entropy. 
When the subsystem is much smaller than the total system, it suffices to
keep the linear order terms. Then, $P_{\rm L}(Q,n)$ reduces to 
the grand canonical ensemble distribution. 
In the current situation, however, the subsystem
size is comparable to the total system size. It is necessary to keep 
the higher order terms.

\begin{figure}
\includegraphics*[width=\columnwidth]{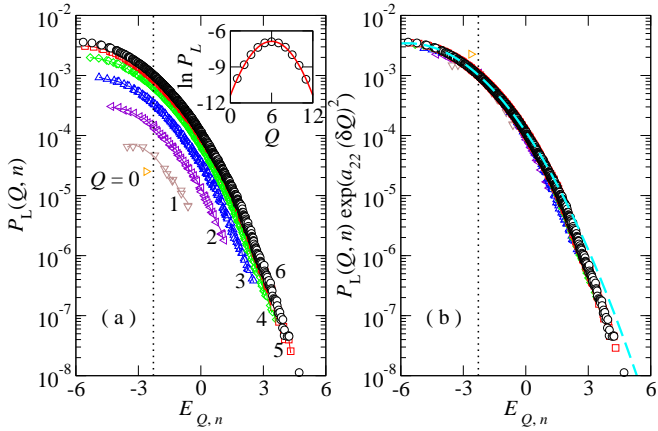}
\caption{(a) Plots of $P_{\rm L}(Q,n)$ with $Q=0,\cdots,6$~(symbols) and 
$Q=7,\cdots,12$~(lines). 
The particle-hole symmetry, $P_{\rm L}(Q,n) = P_{\rm L}(N-Q,n)$, is evident. 
(b) Rescaled probabilities for $Q=0,\cdots,6$. The dashed line is the fitting
curve. 
The vertical dotted lines in (a) and (b) mark the average energy 
$\overline{E_{\rm L}}$.
The inset in (a) plots the probability of the energy level
whose energy eigenvalue is closest to the average energy. 
Parameter values are the same as in Fig.~\ref{fig3}.
}
\label{fig4}
\end{figure}

We confirm the equilibrium distribution in \eqref{P_exp}
numerically. Starting from the same initial condition as in Fig.~\ref{fig3},
we constructed the reduced density matrix $\hat{\rho}_{\rm L}$ averaged over 
the time interval $10^2\leq t \leq 10^3$ when the system reaches the steady
state, and calculated $P_{\rm L}(Q,n)$ using
\eqref{P_def}~(see Fig.~\ref{fig4}).
In the half-filling case~($\overline{Q_{\rm L}}/N = 1/2$), the system 
obeys the particle-hole symmetry.
Thus, the distribution function takes a simpler form 
\begin{equation}\label{P_simple}
P_{\rm L}(Q,n) \propto \exp[ -(\delta E) /T - 
a_{11} (\delta E)^2 - a_{22} (\delta Q)^2] 
\end{equation}
with $\mu=a_{12}=0$. 
In order to verify the quadratic dependence on $\delta Q$, we select the
eigenstate whose energy eigenvalue $E_{Q,n}$ is closest to the average energy
$\overline{E_{\rm L}}$ at each $Q$. The probabilities at these energy levels are
plotted in the inset of Fig.~\ref{fig4}(a). The quadratic dependence is
evident with  $a_{22} \simeq 0.125$.
We plot the rescaled probability distributions $P_L(Q,n)
e^{a_{22} (\delta Q)^2}$ with $Q=0,\cdots,6$ in Fig.~\ref{fig4}(b). 
They collapse onto a single curve, 
which is well fitted to the function 
$\sim e^{-(\delta E)/T - a_{11}(\delta E)^2}$ with $T \simeq 1.40$ and
$a_{11} \simeq 0.105$. The data collapse confirms the equilibrium
distribution function in \eqref{P_simple}.
One can notice a slight deviation at large values 
of $(\delta E)$ and $(\delta Q)$, where even higher order corrections are 
necessary.

{\it Joule expansion --- } 
During the thermalization, the quantum gas undergoes the Joule expansion or
the free expansion into a vacuum. 
We discuss the thermodynamic consequence of the Joule expansion.

Suppose that the system is prepared in the initial state \eqref{Psi_ini} with 
$Q_{\rm L}^0 = 0$ and $Q_{\rm R}^0 = q N$. The subsystem ${\rm R}$ can be in any state with
$n_{\rm R}^0 = 1,\cdots, \binom{N}{qN}$. For an initial state
with given $n_{\rm R}^0$, the gas has a definite initial temperature $T_i$
determined from \eqref{T_E}. 
The final temperature $T_f$ after thermalization can be measured by
fitting the probability distribution $P_{\rm L}(Q,n)$, defined in
\eqref{P_def}, to the form of \eqref{P_exp}. 

\begin{figure}
\includegraphics*[width=\columnwidth]{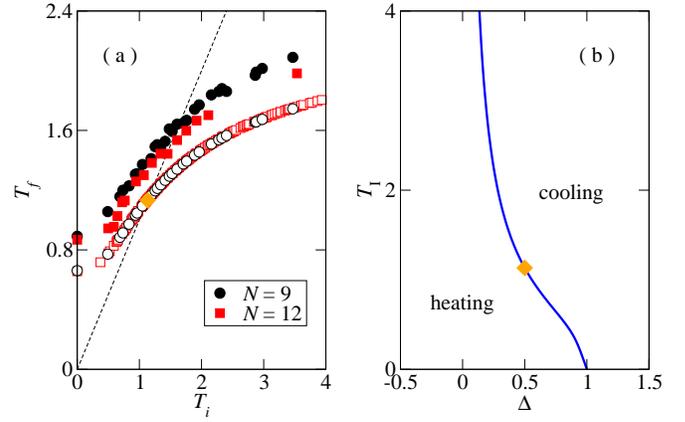}
\caption{(a) Comparison of the temperature before and after the Joule
expansion at $\Delta = 1/2$ and $\lambda=1$. 
The filled symbols represent the temperatures obtained from the
fitting of $P_{\rm L}$, while the open symbols from the isotherm curves.
The dotted line represents the line where $T_f = T_i$.
(b) Inversion curve at $\lambda=1$. 
The inversion temperature at $\Delta = 1/2$ is marked with the symbol
{\color{orange} $\blacklozenge$}.}
\label{fig5}
\end{figure}

We have performed the numerical analysis with the subsystems of size $N=9$
and $12$ and $q=2/3$. 
In fitting, we used the data of the most probable sector 
$Q=\overline{Q_{\rm L}}=(Q_{\rm L}^0+Q_{\rm R}^0)/2$ where the probability takes the simple
form $P_{\rm L} \propto \exp[-(\delta E)/T - a_{11} (\delta E)^2]$ with 
$\delta Q=0$.
The resulting temperature $T_f$ is plotted as a
function of $T_i$ in Fig.~\ref{fig5}(a). 
Interestingly, the quantum gas may either heat up or cool down depending on 
the initial temperature. The two regions are separated by a 
inversion temperature $T_{\rm I}$. We also studied the
Joule expansion from a dense region to a dilute~(nonempty) region 
and obtained the similar result~\cite{Note1}.

Taking it for granted that the system thermalizes, the heating or cooling
can be understood easily. Let $e_\alpha$ and $q_\alpha$ be the initial energy 
density and particle density of the subsystem $\alpha$. 
After the interaction turns on, 
the total system has the energy density $e_f = (e_{\rm L}+e_{\rm R})/2 +
\langle \hat{H}_{\rm int} \rangle_0 / (2N)$ and the particle
density $q_f = (q_{\rm L}+q_{\rm R})/2$. The $O(1/N)$ 
correction to the energy density is negligible in the large $N$ limit. 
One may represent the thermodynamic state of the subsystems and the total
system in the energy density-particle density plane along with the
isotherms~(see Fig.~\ref{fig2}(b)).
In this plane, the total system after expansion is represented by the
midpoint of $(q_{\rm L},e_{\rm L})$ and $(q_{\rm R},e_{\rm R})$. The final temperature $T_f$ can be
read from the isotherm passing through $(q_f,e_f)$.
We illustrate this construction in Fig.~\ref{fig2}(b), where the initial states
are marked as $A$ and $B$ while the final state as $F$. It 
clearly shows that the gas heats up~(cools down) if the isothermal curves
are convex~(concave). We compare the final temperatures from the isotherms
and the probability distributions in Fig.~\ref{fig5}(a). 
A little discrepancy is due to the $O(1/N)$ correction to the energy
density.

At the inversion temperature $T_{\rm I}$, the isotherms change the 
convexity and the $T$-$e$ curves in all $Q$ sectors cross 
each other~(see Fig.~\ref{fig2}(a)). We can locate the inversion
temperature under reasonable assumptions: (i) $T$-$e$
curves in $Q=0$ and $Q=1$ sectors cross at $T_{\rm I}$. (ii) The periodic
boundary condition yields the same result in the $N\to\infty$ limit. 
Under these assumptions, the inversion temperature is determined by
\begin{equation}
-2\Delta = \frac{ \int_{-\pi}^\pi dk\ \varepsilon(k) e^{
- \varepsilon(k)/T_{\rm I}} }{ \int_{-\pi}^\pi dk\ e^{
-\varepsilon(k) / T_{\rm I}}}
\end{equation}
with $\varepsilon(k) = -\frac{2}{(1+\lambda)}(\cos k + \lambda \cos
2k)$~\cite{Note1}.
We evaluate the inversion temperature $T_{\rm I}$ as a function of 
$\Delta$.
The inversion curve $T_{\rm I} = T_{\rm I}(\Delta)$ thus-obtained at $\lambda=1$ 
is presented in Fig.~\ref{fig5}(b).
It vanishes as $T_{\rm I} \sim (\Delta_c-\Delta)$ near $\Delta \simeq \Delta_c = 1$ 
and diverges as $T_{\rm I} \sim \Delta^{-1}$ near $\Delta \simeq 0$.

The inversion curve has an interesting implication. 
Suppose that one can perform the Joule expansion repeatedly. 
Then, the temperature of the gas converges to the inversion temperature 
for $0<\Delta < \Delta_c$. That is, the inversion temperature is the {\em
stable} fixed point under the Joule expansion.

The classical gas with the Lennard-Jones potential also 
has the inversion temperature~\cite{Goussard:1993iy}.
The mean field theory shows that the gas heats up above the inversion
temperature and cools down below the inversion temperature. Namely, the
inversion temperature in the classical regime corresponds to the {\em
unstable} fixed point under the Joule expansion. The distinct role of the
two inversion temperatures manifests that the Joule expansion of the quantum
gas is a correlated many-body process.

{\it Summary --- }
We investigated the equilibration of two subsystems of comparable
size. They are coupled by an interaction Hamiltonian allowing the flows 
of energy and particle.
Based on the ETH, we derived that
the probability distribution of the subsystem follows the grand 
canonical ensemble distribution with nonlinear corrections. 
Our setup describes a Joule expansion of a quantum gas. 
We showed that the quantum gas can be cooled or heated 
upon expansion with the inversion temperature. 
The emergent inversion temperature corresponds to the stable
fixed point under the Joule expansion.
It signals that the Joule expansion of the quantum gas is a correlated
many-body process, which requires a thorough investigation beyond the 
classical mean field theory.
We expect that the cooling/heating scenario can be verified using the free
expansion experiments of ultracold atom gases. We also expect that 
the Joule expansion can be useful for the temperature control of quantum
gases.

\begin{acknowledgments}
This work was supported by the the National Research
Foundation of Korea (NRF) grant funded by the Korea
government (MSIP) (No. 2016R1A2B2013972). 
E.I. and T.S. are supported by JSPS KAKENHI Grand Number JP16H02211. 
E.I. is also supported by KAKENHI Grant Number JP15K20944.
\end{acknowledgments}

\bibliographystyle{apsrev4-1}
\bibliography{preprint}

\widetext

\section{Supplmental material}

\renewcommand{\theequation}{S\arabic{equation}}
\renewcommand{\thefigure}{S\arabic{figure}}
\setcounter{equation}{0}
\setcounter{figure}{0}

\subsection{Numerical method to solve the Schr\"odinger equation}
The state vector at time $t$ is given by 
$|\Psi(t)\rangle = \hat{U}(t) |\Psi(0)\rangle$ with
the unitary time evolution operator 
\begin{equation}
\hat{U}(t) \equiv e^{-i \hat{H}t} .
\end{equation}
When the whole set of eigenvectors $\{|a\rangle\}$ of 
$\hat{H}$ is available, 
it is given by $|\Psi(t)\rangle = \sum_a c_a e^{-i E_a t} 
| a \rangle$ with $c_a = \langle a|\Psi(0)\rangle$. 
This is the exact method.

\begin{figure}[h]
\includegraphics*[width=0.5\columnwidth]{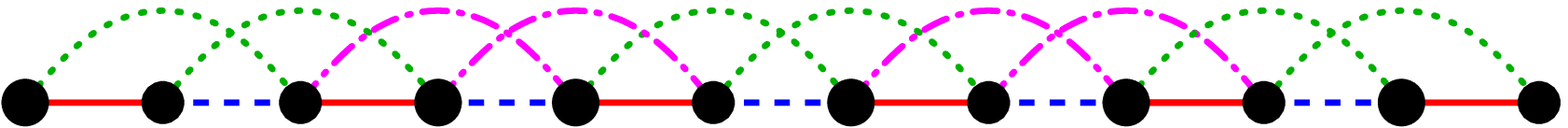}
\caption{Illustration of the XXZ Hamiltonian with the nearest and next
nearest interactions under the open boundary condition.}
\label{sfig1}
\end{figure}

When the dimension of the Hilbert space is too large, the exact method is
impractical and one needs to rely on an approximate method. 
We will explain an efficient method to simulate the time evolution.
The XXZ Hamiltonian of $N$ spins under the open boundary condition 
is given by~(see Fig.~\ref{sfig1})
\begin{equation}
\hat{H} = \sum_{l=1}^{N-1} \hat{h}_{l,l+1} + \sum_{l=1}^{N-2} \hat{h}_{l,l+2} 
\end{equation}
with the short-hand notation
\begin{equation}
\begin{aligned}
\hat{h}_{l,m} \equiv h(\hat{\sigma}_l,\hat{\sigma}_m) &= -\frac{J}{2} \left( \hat{\sigma}_l^x\hat{\sigma}_m^x + 
                                      \hat{\sigma}_l^y\hat{\sigma}_m^y + 
                        \Delta \hat{\sigma}_l^z \hat{\sigma}_m^z \right) \\
  & = -J \left( \hat{\sigma}_l^+ \hat{\sigma}_m^- + 
                \hat{\sigma}_l^- \hat{\sigma}_m^+ 
      + \frac{\Delta}{2} \hat{\sigma}_l^z \hat{\sigma}_m^z \right) ,
\end{aligned}
\end{equation}
where $\hat\sigma_l^{\pm} = (\hat\sigma_l^x \pm i \hat\sigma_l^y)/2$.

Note that the Hamiltonian is slightly different from that considered in the
paper. 
Generalizations to the Hamiltonian with bond-dependent $J$ and $\Delta$ 
are straightforward.

The time evolution operator satisfies $\hat{U}(\epsilon M) =
\hat{U}(\epsilon)^{M}$. Thus, the state vector at $t = \epsilon M$
is obtained by multiplying $\hat{U}(\epsilon)$ to the initial 
state vector $M$ times repeatedly. 
For infinitesimal $\epsilon$, $\hat{U}(\epsilon)$ can be approximated
efficiently.  First, we decompose the Hamiltonian as the sum 
$\hat{H} = \hat{H}_A + \hat{H}_B + \hat{H}_C + \hat{H}_D$ with
\begin{equation}\label{decomp}
\begin{split}
\hat{H}_A & =  \sum_{l} \hat{h}_{2l-1,2l} \\
\hat{H}_B & =  \sum_{l} \hat{h}_{2l,2l+1} \\
\hat{H}_C & =  \sum_{l} ( \hat{h}_{4l-3,4l-1} + \hat{h}_{4l-2,4l})\\
\hat{H}_D & =  \sum_{l} ( \hat{h}_{4l-1,4l+1} + \hat{h}_{4l,4l+2} ) .
\end{split}
\end{equation}
In Fig.~\ref{sfig1}, the interaction bonds contained in 
$\hat{H}_A$, $\hat{H}_B$, $\hat{H}_C$, and $\hat{H}_D$
are represented by the solid, dashed, dotted, and dash-dotted lines,
respectively.
Then, using the generalized Lie-Suzuki-Trotter formula of the second order
$e^{\epsilon(\hat{X}+\hat{Y})} = e^{\frac{\epsilon}{2}\hat{X}} e^{\epsilon
\hat{Y}} e^{\frac{\epsilon}{2}\hat{X}} +
O(\epsilon^3)$~[30], we obtain
\begin{equation}
\hat{U}(\epsilon) =  e^{-i\frac{\epsilon}{2}\hat{H}_A}
                     e^{-i\frac{\epsilon}{2}\hat{H}_B}
                     e^{-i\frac{\epsilon}{2}\hat{H}_C}
                     e^{-{i\epsilon}\hat{H}_D}
                     e^{-i\frac{\epsilon}{2}\hat{H}_C}
                     e^{-i\frac{\epsilon}{2}\hat{H}_B}
                     e^{-i\frac{\epsilon}{2}\hat{H}_A} + O(\epsilon^3)
\end{equation}

The important feature of the decomposition \eqref{decomp} is that all
$\hat{h}_{l,m}$'s contained in a partial Hamiltonian commute with each
others. It yields
\begin{equation}
e^{-i\gamma \hat{H}_A} = \prod_{l} e^{-i \gamma \hat{h}_{2l-1,2l}}
\end{equation}
and similar relations for the others. Consequently, $\hat{U}(\epsilon)$ can
be approximated as the ordered product of $\hat{U}_{l,m}(\gamma) = e^{-i \gamma
\hat{h}_{l,m}}$ with $\gamma=\epsilon$ or $\epsilon/2$ with an error $O(\epsilon^3)$.
If one chooses
$\{ |-\rangle_l\otimes |-\rangle_m, 
|-\rangle_l\otimes |+\rangle_m, 
|+\rangle_l\otimes |-\rangle_m,
|+\rangle_l\otimes |+\rangle_m \}$ as the basis states for the spins at
sites $l$ and $m$, $\hat{U}_{l,m}(\gamma) = \hat{u}_{l,m}(\gamma) \otimes
\mathsf{I}_{N-2}$ with the $4\times 4$ unitary matrix
\begin{equation}
\hat{u}_{l,m}(\gamma) = 
\begin{pmatrix}
e^{i \phi } & 0 & 0 & 0 \\
0 & e^{-i\phi} \cos\gamma J& - i e^{-i\phi} \sin \gamma J & 0 \\
0 & -i e^{-i\phi} \sin\gamma J&  e^{-i\phi} \cos \gamma J & 0  \\
0 & 0 & 0 & e^{i \phi} 
\end{pmatrix}
\end{equation}
with $\phi = \gamma J \Delta /2$.
The overall error of this method is of the order of $O(\epsilon^2)$.

\begin{figure}
\includegraphics*[width=0.5\columnwidth]{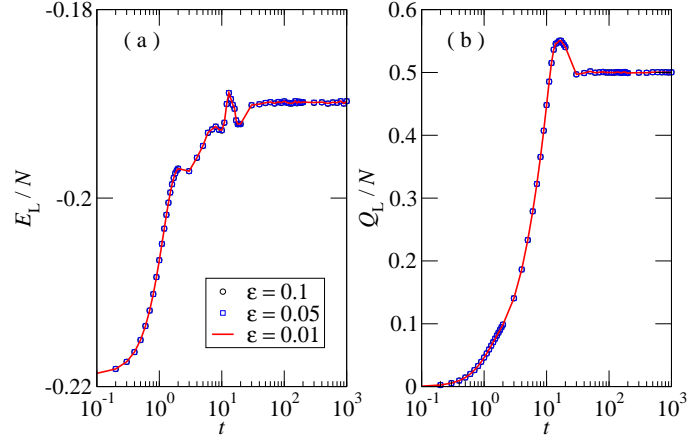}
\caption{Numerical data for the energy density and the particle density for
the subsystem L using the same condition as Fig.~3 of the main text. The
data obtained by using $\epsilon=0.1, 0.05,$ and $0.01$ cannot be
distinguished.}
\label{sfig2}
\end{figure}

Computationally, the infinitesimal time evolution is achieved by the
multiplication of a state vector with the $4\times 4$ matrices $O(N)$ times.
All the operators $\hat{U}_{l,m}$'s conserve the total number of up spins. 
Thus, one can work within the subspace consistent with the conservation,
which make numerical calculations efficient.
In the numerical calculations, we used the exact time evolution when $N\leq
16$, and the decomposition method for $N>16$ with $\epsilon = 0.01$. We
examined whether $\epsilon = 0.01$ is small enough. As can be seen in
Fig.~\ref{sfig2}, the data with $\epsilon=0.1, 0.05$, and $0.01$ are 
indistinguishable in the plot. More quantitatively, the values of 
$Q_{\rm L}$ at time $t=10^3$ are $6.004072339$, $6.003279566$, and 
$6.003025718$ for $\epsilon=$ $0.1$, $0.05$, and $0.01$, respectively. 
They are well fitted to the function $Q_{\rm L} = Q_0 + a \epsilon^{1.99926}$ with
$Q_0 = 6.003015128$. The relative error of the value at $\epsilon=0.01$ is
$\simeq 1.7\times 10^{-6}$. It proves that $\epsilon=0.01$ is small enough.

\subsection{Singular value decomposition for reduced density matrix}
Suppose that one separates the $N$ spins into two sets $A$ of $N_A$ spins 
and $B$ of $N_B = N-N_A$ spins. We assume that $N_A \leq N_B$ without loss
of generality.
Then, the state vector can be written as
\begin{equation}
|\Psi\rangle = \sum_{a,b} \psi_{a,b} |a\rangle\otimes |b\rangle ,
\end{equation}
where $\{|a\rangle\}$ and $\{|b\rangle\}$ span the Hilbert spaces of
dimension $d_A=2^{N_A}$ and $d_B = 2^{N_B}$ for spins in $A$ and $B$, 
respectively. The reduced density operator for the spins in $A$ is
given by 
\begin{equation}
\hat{\rho}_A = {\rm Tr}_B |\Psi\rangle\langle \Psi| = \sum_{b}
\psi_{a,b}\psi_{a',b}^* |a\rangle \langle a'| 
\end{equation}
and represented by the $({d_A}\times {d_A})$ reduced density matrix 
is given by
\begin{equation}
(\rho_A )_{a,a'} = \sum_b \psi_{a,b} \psi_{a',b}^* .
\end{equation}
The reduced density matrix is necessary when one calculates the local 
observables and the probability distribution for the spins in $A$. 
The eigenvalues are necessary
for the entanglement entropy. The singular value decomposition~(SVD) is a
convenient method to handle the reduced density matrix.

The probability amplitudes $\psi_{a,b}$ can be regarded as the 
elements of a $({d_A}\times {d_B})$ positive semidefinite complex matrix 
$M$, $M_{ab} =
\psi_{a,b}$. According to the SVD, such a matrix $M$ can be written as
\begin{equation}
M = V_1 \Sigma V_2^\dagger , 
\end{equation}
where $V_1$ is a $({d_A}\times {d_A})$ unitary matrix, $\Sigma$ is a $({d_A}
\times {d_B})$ rectangular diagonal matrix with non-negative real diagonal
elements, and $V_2$ is a $({d_B}\times {d_B})$ unitary matrix. 
The SVD can be done with a numerical analysis software. 

Once the SVD is done, the reduced density matrix is obtained from the
relation
\begin{equation}
\rho_A = M M^\dagger = V_1 \Sigma V_2^\dagger V_2 \Sigma^\dagger V_1^\dagger
       = V_1 \Sigma \Sigma^\dagger V_1^\dagger .
\end{equation}
That is, the reduced density matrix is given by the similarity transformation 
of the diagonal $(d_A\times d_A)$ matrix
\begin{equation}
\Sigma \Sigma^\dagger = {\rm
diag}(|\lambda_1|^2,|\lambda_2|^2,\cdots,|\lambda_{d_A}|^2) .
\end{equation}
The SVD allows one to calculate the entanglement entropy easily. 
It is given by
\begin{equation}
S_{ent} = -\sum_{n=1}^{d_A} |\lambda_n|^2 \ln |\lambda_n|^2 .
\end{equation}

\subsection{Temperature-energy relation in the $Q=1$ sector}
Consider the XXZ Hamiltonian of $N$ spins under the periodic boundary
condition. In the $Q=0$ sector where all spins are down~(or all sites are
empty), there is a single eigenstate with the energy 
\begin{equation}
E_{0,1} = -\frac{\Delta}{2} N .
\end{equation}
The Hilbert space of the $Q=1$ sector is spanned by the states $\{|1\rangle,
\cdots, |x\rangle,\cdots,|N\rangle\}$ where $|x\rangle$ represents the state
where the particle is at site $x$. The plane
wave $|k_n\rangle = \frac{1}{\sqrt{N}} \sum_x e^{ik_n x} |x\rangle$
is the eigenstate and the energy eigenvalue is given by
\begin{equation}
E_{1,k_n} = -\frac{\Delta}{2}(N-4) - \frac{2}{1+\lambda} \left( \cos k_n +
\lambda \cos 2k_n \right) = -\frac{\Delta}{2}(N-4) +
\varepsilon(k_n)
\end{equation}
with $k_n = 2\pi n / N$~($n=0,\cdots,N-1$). The partition function in the
$Q=1$ sector is given by
\begin{equation}
Z_1(\beta) = 2^{\beta \Delta (N-4)/2}\sum_{n=0}^{N-1}  e^{-\beta
\varepsilon(k_n)} 
\end{equation}
with the mean energy
\begin{equation}
\langle E\rangle = - \frac{\partial}{\partial \beta} \ln Z_1(\beta)
= -\frac{\Delta}{2}(N-4) + \frac{\sum_n \varepsilon(k_n) e^{-\beta
\varepsilon(k_n)}}{\sum_n e^{-\beta \varepsilon(k_n)}} .
\end{equation}

We estimate the location of the inversion temperature $T_{\rm I}$ 
by requiring that
the mean energy in the $Q=1$ sector should be equal to the energy in the $Q=0$
section. It yields that
\begin{equation}
-2 \Delta =  \frac{\sum_n \varepsilon(k_n) e^{-\beta_{\rm I}
\varepsilon(k_n)}}{\sum_n e^{-\beta_{\rm I} \varepsilon(k_n)}} 
\end{equation}
with $\beta_{\rm I} = 1/T_{\rm I}$. In the $N\to\infty$ limit, the summation is replaced
by the integration to yield the expression
\begin{equation}
-2\Delta = \frac{ \int_{-\pi}^\pi dk \ \varepsilon(k) e^{-\beta_{\rm I}
\varepsilon(k)}}{ \int_{-\pi}^\pi dk \ e^{-\beta_{\rm I} \varepsilon(k)}}  \equiv
\langle \varepsilon(k)\rangle_{\beta_{\rm I}}.
\end{equation}

Note that $\langle \varepsilon(k)\rangle_{\beta_{\rm I}}$ increases as
$\beta_{\rm I}
\to 0^+$ or $T_{\rm I} \to +\infty$. For small $\beta_{\rm I}$, we can approximate
\begin{equation}
\langle \varepsilon(k)\rangle_{\beta_{\rm I}} \simeq -\beta_{\rm I} \int \frac{dk}{2\pi}
\varepsilon(k)^2  = -2 \beta_{\rm I} \frac{1+\lambda^2}{(1+\lambda)^2} . 
\end{equation}
Thus, the inversion temperature has the limit form
\begin{equation}
T_{\rm I} \simeq \frac{1+\lambda^2}{(1+\lambda)^2} \Delta^{-1}
\end{equation}
as $\Delta \to 0^{+}$.
In the opposite limit where $\beta_{\rm I} \to \infty$~($T_{\rm I}\to 0^+$), $\langle
\varepsilon(k)\rangle_{\beta_{\rm I}}$ is dominated by the minimum value of
$\varepsilon(k)$. For positive $\lambda$, $\varepsilon(k)$ takes the minimum
value $-2$ at $k=0$. It yields that $T_{\rm I} \simeq (1-\Delta)/4$.

\subsection{Joule expansion}
In the main text, we present the Joule expansion of a gas into the
vacuum. We have performed the numerical simulation of the Joule expansion
in a mixture of a dense gas and a dilute gas.
The subsystems ${\rm L}$ and ${\rm R}$ is prepared in the energy eigenstates
$|Q_{\rm L}^0 =
q_{\rm L} N, n_{\rm L}\rangle$ and $|Q_{\rm R}^0 = q_{\rm R} N, n_{\rm R}\rangle $, respectively, 
with $q_{\rm L} = q = 1/4$ and $q_{\rm R} = 1-q = 3/4$ with the same quantum
number $n_{\rm L}
= n_{\rm R} = n_0$. 
The two states are tied to each other by the particle-hole symmetry.
We select those two states in order to keep them in the same initial
temperature.

\begin{figure}
\includegraphics*[width=0.5\columnwidth]{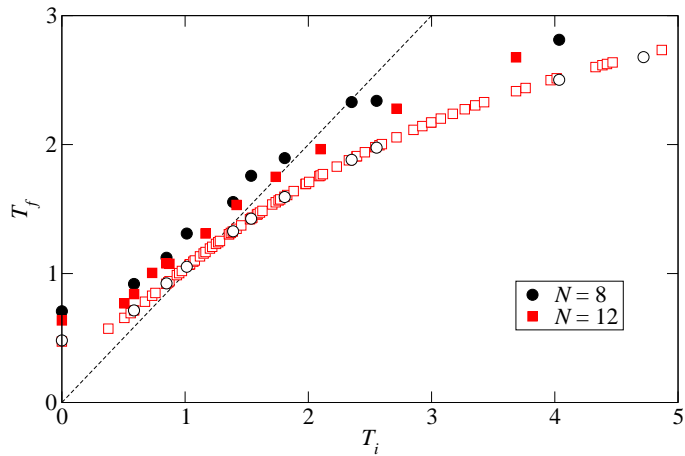}
\caption{Heating/cooling curve when the dense gas of density $\rho_{\rm R}=3/4$ is
mixed with the dilute gas of density $\rho_{\rm L} = 1/4$. The filled symbols
represent the numerical results obtained by fitting the probability 
distribution. The open symbols represent the expected result from 
the isotherm curve.}
\label{sfig3}
\end{figure}

When the total system reaches a stationary state, we measure the probability
distribution $P_{\rm L}(Q,n)$ that the subsystem ${\rm L}$ is in the state
$|Q,n\rangle$. Then, we fit the probability distribution to the function
$P_{\rm L}(Q,n) \propto \exp[ -(E_{Q,n}-\overline{E_{\rm L}})/T_f - a_{11}
(E_{Q,n}-\overline{E_{\rm L}})^2]$ at the most probable sector $Q=N/2$ to find the
final temperature $T_f$. The numerical result is presented in
Fig.~\ref{sfig3}. We also observe that the gas heats up (cools down) when
the initial temperature is lower (higher) than a threshold temperature.
The finite size effect is also seen in the plot.

\end{document}